\documentclass[11pt]{article}
\usepackage{amsfonts}
\usepackage{graphicx}
\usepackage{amsmath}
\makeatletter \@addtoreset{equation}{section}

\newcommand{\be}{\begin{equation}}
\newcommand{\ee}{\end{equation}}
\newcommand{\bea}{\begin{eqnarray}}
\newcommand{\eea}{\end{eqnarray}}
\begin{document}
\title{%
\begin{flushright}
 {\normalsize \small
GNPHE/10-03 }
 \\[.5cm]
 \mbox{}
\end{flushright}
\textbf{ Engineering of Quantum Hall Effect from Type IIA String
Theory on The K3 Surface } }
\author{ \hspace*{-20pt}
 Adil Belhaj$^{1,3}$\thanks{\tt{belhaj@unizar.es}}, Antonio Segui$^{2}$\thanks{\tt{segui@unizar.es}},
\\
%EndAName
\\ {\small $^{1}$Centre National
de l'Energie, des Sciences et des Techniques Nucleaires, CNESTEN }\\
[-6pt] {\small  Rabat, Morocco } \\{\small $^{2}$Departamento de
F\'isica Te\'orica, Universidad de
Zaragoza, E-50009-Zaragoza, Spain}\\
{\small $^{3}$Groupement National de Physique des Hautes Energies, GNPHE, Si\`{e}ge focal: FSR}\\
[-6pt] {\small  Rabat, Morocco } } \maketitle
\begin{abstract}
Using   D-brane  configurations  on the K3 surface,  we give six
dimensional type IIA stringy realizations of the  Quantum Hall
Effect (QHE)
 in 1+2 dimensions.  Based    on the vertical and horizontal
lines of the K3 Hodge diamond, we  engineer two different stringy
realizations. The  vertical  line  presents  a realization in terms
of D2 and D6-branes wrapping the K3 surface. The horizontal one  is
associated with  hierarchical  stringy descriptions obtained from a
quiver gauge theory  living  on a stack of  D4-branes wrapping
intersecting 2-spheres  embedded in the  K3 surface with
deformed singularities. These geometries   are classified by three kinds of the
Kac-Moody algebras: ordinary, i.e finite dimensional, affine and
indefinite. We find that no stringy QHE in 1+2 dimensions can occur
in the quiver gauge theory living  on intersecting  2-spheres arranged as  affine Dynkin diagrams. Stringy realizations of QHE  can be done
only for the finite and indefinite geometries. In particular, the
finite Lie algebras give models  with fractional filling fractions,
while
  the indefinite ones  classify    models with negative  filling
 fractions  which can be  associated  with the physics of  holes in the graphene.
 \newline
\textbf{Keywords}: Quantum Hall Effect, Type IIA  superstring, K3
surface,  Quiver  gauge models.
\end{abstract}
%\newpage
%\tableofcontents
\thispagestyle{empty}
\newpage \setcounter{page}{1} \newpage

\section{Introduction}
Susskind  has conjectured that a two-dimensional quantum Hall fluid
of charged particles with filling fraction $\nu=\frac{1}{k}$  is
described by a non-commutative Chern-Simons  gauge theory at level
$k$ \cite{S}. This conjecture  has opened a  new way  to apply
string theory  to study low-energy systems in condensed matter
physics  as the  Quantum Hall Effect (QHE)  in $1+2$ dimensions.
This has been based on the recent developments in string dualities
and   brane physics. Roughly, the first connection with string
theory  has been given by Bernevig, Brodie, Susskind and Toumbas to
reproduce the QHE on the 2-sphere, $S^2$ \cite{BBST}. The
corresponding brane configuration involves a spherical D2-brane and
dissolves D0-branes on it. The system  has been placed in a
background of coincident D6-branes extended in the directions
perpendicular to the world-volume of the D2-brane on which QHE
resides. Some further details on  this construction can be found in
\cite{BN}-\cite{AARS}. However, this type IIA superstring construction
has been done in ten dimensional space-time. Therefore a natural
 question arises whether the  QHE in 1+2 dimensions can
be realized in terms of lower dimensional  type IIA superstring
spectrums.

The aim of this work  is to  address this question  by giving
stringy realizations of the  QHE in $1+2$ dimensions from a six
dimensional string  space-time  point of  view. Motivated by the
paper \cite{FLRT} and the results obtained on the attractor horizon
geometries of the extremal black 2-branes \cite{BDSS,AABS}, we
engineer brane representations of the  QHE in $1+2$ embedded in  the
compactification of type IIA superstring  on the K3 surface. Based
on the vertical and horizontal lines of the Hodge diamond of the  K3
surface, we give two different  stringy realizations. The vertical
line is associated with a realization in terms of D2 and D6-branes
wrapping the K3 surface. However, the horizontal line corresponds to
a quiver gauge theory realization of the  QHE in 1+2 dimensions.
This gives hierarchical stringy  descriptions in terms of  wrapped
D4-branes on intersecting  2-spheres of  local  limits of the K3
surface. The intersection numbers are encoded
in  the
   Cartan
matrices classified by three kinds of the Kac-Moody algebras
\cite{V,ABS,AABDS}. They are known as: ordinary i.e  finite
dimensional,
 affine and  indefinite.  We find that no stringy QHE in 1+2
dimensions can occur in the quiver gauge theory living on a stack of
D4-branes
 based on  the  affine Dynkin diagrams.   However,  the finite Lie algebras  classify
 models with fractional filling fractions; while
  the indefinite ones   give   models with negative  filling
 fractions, which could   be associated  with the holes in the graphene physics.

The organization of this paper is as follows. In section 2 we review
 briefly the stringy realization of the  QHE in $1+2$ dimensions   in terms of ten dimensional  type  IIA
superstring. In section 3,  we  give  a six dimensional   stringy realizations of the
QHE in $1+2$ dimensions  from  the compactification  of type IIA
superstring on the K3 surface. Based on the Hodge diamond vertical
line of the K3 surface, we give a realization in terms of D2 and
D6-branes wrapping the K3 surface. In section 4, we present
hierarchical  stringy descriptions obtained from quiver gauge
theories living on the world-volume of D4-branes wrapping intersecting 2-spheres of
local limits of the  K3 surfac with  geometries classified bt the three kinds of Kac-Moody Lie algebras. These realizations, which are based on the horizontal
line,  lead to the QHE with filling fractions depending on local  type IIA
 geometries. The last section is devoted to discussions and
open questions.
\section{ Uncompactified String Theory Realization  of the QHE}
 Following \cite{BBST},
3-dimensional QHE systems of condensed matter physics   can be
modeled by low energy dynamics of non perturbative extended objects,
involving D0, D2 and D6-branes of  uncompactified type IIA
superstring living  in ten dimensions. To see that, consider  first
a D2-brane, with  a spherical configuration, and dissolves $N$
D0-branes moving on it. Second we take a stack of D6-branes  with
$K$ charges placed in the directions perpendicular to the ones where
the  spherical D2-brane lives.  Then we   move them to the center of
the 2-sphere. In this brane configuration,  the world-volume of the
D2-brane plays the role of the space-time of the 3-dimensional QHE
systems. It can be also interpreted as the space-time of the
3-dimensional Chern Simons theory with $U(1)$  gauge group living on
the D2-brane.  The D6-branes, located at the origin of the D2-brane,
 can be thought of as an external source of the  magnetic charges. This
 has been required   from the fact that   in ten dimension  a
 $p$-dimensional   electrical   brane is  dual to a
$q$-dimensional magnetic one  such that
\begin{equation}
\label{gd} p+q=6.
\end{equation}
When the D6-branes cross the D2-brane, the Hanany-Witten effect
produces fundamental strings  which  are stretched  between D2 and
D6-branes \cite{HW}. In this way, the strings ending on the D2-brane
have an interpretation in terms of the fractional quantum Hall
particles (Hall electrons). The D0-branes behave as a magnetic flux
in the world-volume of the D2-brane and the string ends are charged
under the  U(1) world-volume gauge field. The number of the
D6-branes $K$ can be identified with the number of charged particles
and $N$ is the total magnetic flux.  For this brane configuration,
the ratio
\begin{equation}
\nu=\frac{K}{N}
\end{equation}
determines  the filling  fraction  of the  system.  \\ At this level,
one might naturally ask the following question. Is there any
compactified string theory realization of the  QHE in 1+2
dimensions?. In what follows,  we address this question using the
compactification  of type IIA superstring   on the K3 surface and
wrapped D-branes on  appropriate cycles.
\section{  Six Dimensional Type  IIA String Realization of the QHE}
Using the   results of \cite{FLRT} and  the   black brane attractors
on the  K3 surface  \cite{BDSS,AABS}, we engineer  six dimensional string
realizations of the  QHE in $1+2$ dimensions.  This can be embedded
in  the compactification of type IIA superstring    on  the K3
surface. This geometry  is a two dimensional complex   Calabi-Yau
manifold admitting a $SU(2)$ holonomy group. It has many types of
realizations; the simplest one is to consider the orbifold $T^4/G$
where  $ G$  is a subgroup of $SU(2)$. In the  case of $G = Z_2$,
the K3 surface is obtained by blowing up the sixteen singular fixed
points of the corresponding orbifold. Each fixed  point  is replaced
by a
 2-sphere $S^2$. The K3 surface involves   a Hodge diamond
 playing a
crucial role in the determination of  the six dimensional string
theory spectrum and the corresponding  brane configurations.  For a
generic  K3 surface,  the Hodge diamond  is given by
 \[
\begin{tabular}{lllll}
&  & $h^{0,0}=1$ &  &  \\
& $h^{1,0}=0$ &  & $h^{0,1}=0$ &  \\
$h^{2,0}=1$ &  & $h^{1,1}=20$ &  & $h^{0,2}=1$. \\
& $h^{2,1}=0$ &  & $h^{1,2}=0$ &  \\
&  & $h^{2,2}=1$ &  &
\end{tabular}%
\]
This graph  shows that  $H_2(K3)$  has dimension 22, corresponding
to the fact that $h^{2,0} = h^{0,2}= 1$ and $h^{1,1}=20$.
 The compactification of type IIA  superstring on the K3 surface is obtained
by breaking the space-time symmetry $SO (1, 9)$ down to the subgroup
$ SO(1, 5)\times SU (2)$  which is contained in $SO (1, 5)\times
SO(4)$. More details on  this compactification can be found  in
\cite{As}.

Similarly as for the  ten dimensional realization of the QHE in
$1+2$ dimensions, the space-time  is described  by the world-volume
of membranes  embedded in   the resulting six dimensional space-time
obtained from the compactification of type IIA superstring on   the
K3 surface. In this compactification, we can produce such membranes
by considering a D-brane system consisting of:\\
\begin{itemize}
\item Unwrapped  D2-branes
\item D4-branes wrapping  2-spheres   inside  the K3 surface
\item   D6-branes wrapping  the K3  surface.
\end{itemize}
After  deleting  the zeros of the Hodge diamond  of the K3 surface,
these 2-branes   can be represented  by the following Hodge brane
configurations\footnote{
 Note in passing  that these brane configurations  has been used in the
 study of attractor horizon  geometries  of extremal black 2-branes
\cite{BDSS}. In such  limits,   type IIA backgrounds can be
identified  with the factorization $AdS_4 \times  S^2\times K3$.}
\def\m#1{\makebox[10pt]{$#1$}}
\begin{equation}
\label{hodgebrane}
  {\arraycolsep=2pt
  \begin{array}{*{5}{c}}
    && D2&& \\ &&&& \\ D4&&D4&&D4.\\
    &&&& \\ &&D6&&
  \end{array}}
\end{equation}
In what follows, we  shall show that  the compactification of  type
IIA superstring on the K3 surface provides a natural realization of
the QHE in 1+2  dimensions. From (\ref{hodgebrane}), we can engineer
at least  two type IIA stringy realizations of this effect. The
vertical  line of the previous Hodge sub-diamond represents a brane
system involving D2 and D6-branes wrapping the K3 surface, while the
horizontal one
 considers only  D4-branes wrapping  intersecting 2-spheres.  Here  we give
 the vertical representation and leave the horizontal one for the
 next section.  Roughly, the   vertical QHE stringy  realization
 can be done in a similar manner as   the
  uncompactified stringy one  discussed previously.  To see that consider an
  unwrapped  D2-brane
   which can be obtained from the ten dimensional
  one. As in the uncompactified case,  its world-volume will  play   the role of
  the 3-dimensional  space-time on which QHE  may reside.  However,   the   magnetic
  source in  six dimensions
  should  have only two extended directions.  This would  be obtained by compactifying the ten
   dimensional brane objects on cycles of the K3 surface. The presence of the membrane magnetic source
  is due to the  fact that the
electric/magnetic duality in six dimensions reads  as
\begin{equation}
p+q=2.
\end{equation}
In the compactification on the K3 surface, the  magnetic  membranes,
dual to electrical particles in six dimensions, are obtained by two
different ways. They are formed by wrapping D4-branes on the
2-spheres of the K3 surface, or by wrapping D6-branes on the entire
K3 surface, and they  lead to a source of the dual 4-form field
strength. These brane objects  carry magnetic charges of the $U(1)$
gauge symmetry obtained from the reduction of the corresponding
higher dimensional {\bf R-R} gauge fields.

In this section
we restrict our-self to the case of the  wrapped D6-branes to form a
magnetic source and leave the D4-brane case  for the next section.
Indeed, consider $N$ coincident of D6-branes wrapping the K3
surface. Wrapping a D6-brane on the K3 surface means that  four
spatial directions are wrapped on the K3 surface, while the
remaining two directions stay unwrapped in six dimensions.  As in
ten dimensions, we need to turn on fundamental strings connecting
these wrapped D6-branes with the D2-brane filling the  3-dimensional
space-time on which
 the QHE  lives. The end of the fundamental strings stretched   on the D2-brane can
be considered as 3-dimensional electrons. We  expect to have a
realization similar to the one given in ten dimensions. This can be
obtained by replacing the unwrapped D6-branes by wrapped D6-branes
on the K3 surface. Since the K3 surface is not flat, there are some
curvature corrections to the brane dynamics which should be taken
into account. In fact,  it has been shown in \cite{JPP}  that the
wrapping of a D6-brane on the K3 surface induces a D2-brane with
negative charge within the uncompactified part. The later is not an
anti D2-brane since it preserves the same supersymmetries that a
D2-brane with that orientation. However, we  can add extra D2-branes
to the unwrapped part of the D6-branes with the  charge number $M$.
In this way, the net D2-brane charge as measured at infinity is
therefore $M-K$. Assuming that this charge number  is positive,  the
filling fraction  can be given by the following ratio
\begin{equation}
\nu=\frac{M-K}{N}.\end{equation} If we take $K=N$, this can be
reduced to
\begin{equation}
\nu=k-1,\qquad k=\frac{M}{N}.
\end{equation}

In order to be able to reach the plateau in which the conductivity
is quantized we need a strong enough magnetic field. In our stringy
construction this means that the physical distance between the D2-branes and the D6 branes must be small.
The fundamental string is streched having one end on the D6 stack and the other end on the D2 stack.
If the distance between both stacks is small, the length of the fundamental string is correspondingly small
and has quantum vibrations. On the other hand if the distance between both stacks is large
(compared with the string length) the string is strongly strched and develops a classical configuration.

Parity is broken on the 2+1 dimensional space-time developed on the D2 branes by the presence of the
strong magnetic field orthogonal to the space. This field is sourced by the magnetic D6-branes. As in \cite{FLRT},
another way to break parity is to add a new stack of D4-branes wrapping two-cycles on the K3 surface.

Finally, the model presented  in this section could be related to other
brane realization given recently in \cite{FLRT} based on D3/D7
systems and using the holographic principle. We expect this can be done by
T-dualizing along the direction in which the fundamental string lives. In fact if we
consider this direction as a compact one, this duality exchanges the type IIA superstring with
the type IIB. Moreover, the spacial dimension of the D-brane should
change by an odd number because under T-duality the Dirichlet and
Neumann boundary conditions are exchanged. After a T-duality on a
circle, which can be identified with the fundamental string direction,
the D2/D6 system can be converted into the D3/D7 one. On the other hand, string-string duality
relating type IIA superstring compactified on the K3 surface with the heterotic
superstring compactified on $T^4$ should also give dual versions of the model presented here.
We believe that these facts deserves to be studied further.

\section{  Hierarchical  Type IIA Stringy   Description of the QHE}
So  far, we  have discussed a QHE description with a single gauge
group. Here, we give    stringy realizations involving multiple
$U(1)$ gauge symmetry. This is known by the hierarchical description
\cite{H}. Apriori there  are many ways to embed QHE abelain gauge
theory with $U(1)^n$ gauge group in type IIA superstring
compactification. A way, which has been discussed  in \cite{FLRT},
consists of getting abelian gauge fields  from the reduction of the
{\bf R-R} 3-form  on 2-spheres of the K3 surface. However, the
method   presented  here is somewhat different. It is based on the
study of quiver gauge theories living on the world-volume of the
wrapped D4-branes. To do so, consider a local IIA geometry where the
K3 surface develops
 singularities classified by Kac-Moody  Lie algebras. In this limit, the K3  surface can be
replaced by  ALE spaces with     $ADE$ singularities. The
deformation of such geometries consists on blowing up the
singularity in question by a collection of intersecting  2-spheres
according to the $ADE$ Dynkin diagrams. In this way,  the root
system $\{\alpha_i\}$ of the $ADE$ Lie algebras can play a crucial
role in the understanding of the K3  algebraic geometry and  the
corresponding quiver gauge theories embedded in type IIA superstring
compactification.  It follows that, there is a one to one
correspondence between the
 $ADE$ Lie  algebras and the $ADE$  geometries  of  local limits of the  K3 surface.  More
precisely, to each  root $\alpha_i$ (say the $ su (n)$  simple root
$\alpha_i= e_i-e_{i+1}$) corresponds a 2-sphere $S^2_i$ of the $ADE$
geometries of  the ALE spaces. The connection is supported by  the
following 2-sphere intersection  relation
\begin{equation}
S^2_i.S^2_j=-\alpha_i. \alpha_j=-K_{ij}
\end{equation}
where $K_{ij}$ is exactly the Cartan matrix of the $ADE$  Lie
algebras.

Roughly, the abelian $U(1)^n$ gauge symmetry in three dimensions,
that we consider here, can be obtained by taking a low energy limit
of wrapped D4-branes on  intersecting   geometries.  To get
that, consider  $N$ D4-branes embedded in type IIA superstring. On
theirs unwrapped world-volume lives a 5-dimensional gauge theory
with $U(N)$ symmetry. In type IIA superstring  moving on the
deformed $ADE$ backgrounds, the D4-branes can be partitioned into
$N_i$ D4-branes wrapping  intersecting 2-spheres  arranged as
Dynkin diagrams. These wrapped branes fill the 3-dimensional
space-time on which
 type IIA QHE  can  reside. In this way, the $U(N)$   gauge group breaks down
 to $\prod_{i=1}^n U(N_i)$ where $n$ is
the number of the  intersecting 2-spheres. Now, identifying $n$ with
the number of D4-branes, the gauge group   living on the unwrapped
part of  the world-volume of D4-branes  can be totally broken to
 $U(1)^n$ gauge symmetry. This can  be obtained by thinking that each single D4-brane
 wraps only one 2-sphere inside  the K3 surface.\\
As in \cite{FLRT}, to couple the system to an external gauge field,
we should add an extra  D4-brane wrapping a particular  2-cycle
playing the role of a magnetic source.  As before, this wrapped
D4-brane  is required by the fact that the total space-time of
uncompactified string theory is  six. Its  unwrapped directions
should be placed perpendicularly to the uncompactified part of the
word-volume of D4-branes on which the QHE resides.    As in the
previous realization, the missing spatial  dimension can be
identified with the fundamental string connecting this new  wrapped
D4-brane with the ones  on which the $U(1)^n$ gauge
theory lives. The full brane system can be described  by the
$U(1)^n$ Chern Simon  gauge theory with the following action
\begin{equation} \label{Chern}
S=\frac{1}{4\pi}\int\sum_{i,j}K_{ij}A^i\wedge
dA^j+\sum_{i}q_i{\tilde A}\wedge dA^i.
\end{equation}
In this action, ${\tilde A}$  is an  external $U(1)$   gauge field
and  $A^i$  are dynamical gauge fields.  $K_{ij}$ is a  $n \times n$
matrix  which  can be related to    the intersection matrix of
2-spheres within  the K3 surface \cite{FLRT}.   $q_i$ is the
electromagnetic charge   carried by  the extra D4-brane wrapping  a
special 2-cycle described by a  linear combination of $S^2_i$
\begin{equation}
[C_2]=\sum_iq_i[S^2_i],
\end{equation}
where $S^2_i$ denote  a basis  of $H_2(K3, Z)$. Following Wen-Zee
mode \cite{WZ},  $K_{ij}$ and $q_i$ are interpreted  as order
parameters classifying the various FQH states and  are related to
the filling fraction  by
\begin{equation}
\label{factor} \nu=q_iK_{ij}^{-1}q_j.
\end{equation}
 From this quantity, we clearly see that  the filling fraction  is related to the geometric data  of type IIA  local
backgrounds
 through the intersection  numbers of the 2-spheres.  Up to some details, these intersection numbers  are classified by the  Cartan
matrices. Following \cite{V}, the possible forms of  $K_{ij}$ may be
grouped basically into three categories. In the language of the
Kac-Moody algebras, these correspond to: \\1. Cartan matrices of the
finite dimensional Lie algebras satisfying $det\; K> 0 $\\ 2. Cartan
matrices for the affine Lie  algebras  with the condition
 $ det \;K = 0$
 \\3. Cartan matrices for the indefinite Lie  algebras where  $ det\; K <
 0$.\\
In terms of   Dynkin diagrams of  the  trivalent  $T_{p,q,r}$
extended Lie algebras,  $ det\; K$ can have a nice expression. Note
that the Dynkin diagram of the $T_{p,q,r}$ Lie algebras involves
three  ordinary $A_p$, $A_q$ and $A_r$ Dynkin chains glued at a
trivalent vertex \vspace{1cm}
\begin{figure}[tbph]
\centering
\begin{center}
%{\figure[image arbres.eps]}
\hspace{1.5cm} \includegraphics[width=12cm]{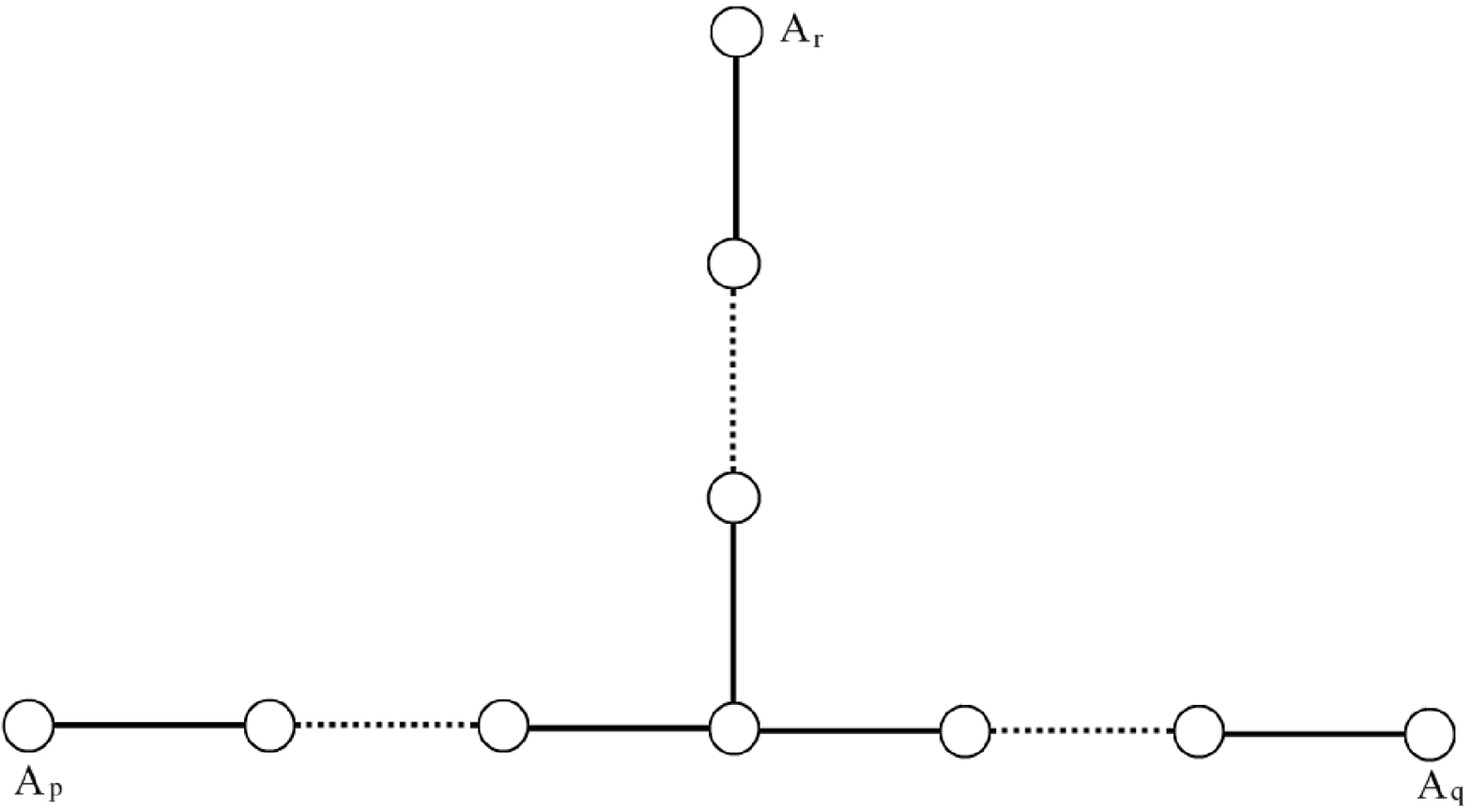}
\end{center}
\end{figure}

The  determinant of the Cartan matrix of the corresponding   Lie
algebra  takes the form
 \begin{equation}
det\;K(T_{p,q,r})=pqr\Gamma
\end{equation}
 where $\Gamma$ is  given by the following quantity
\begin{equation}
\Gamma=\frac{1}{p}+ \frac{1}{q}+\frac{1}{r}-1.
\end{equation}
In this way, the above classification can be reformulated  as
follows: \\1. Cartan matrices of  the  finite dimensional Kac-Moody Lie
algebras satisfying $\Gamma> 0 $\\ 2. Cartan matrices for the affine
Kac-Moody Lie  algebras are given by the  condition
 $ \Gamma = 0$\\
 3. Cartan matrices for indefinite
Kac-Moody  Lie algebras are defined by $ \Gamma < 0$.\\

From  both classifications, it follows  that no QHE in 1+2
dimensions can occur  in the quiver gauge theory of D4-branes
wrapping  on intersecting  2-spheres according to the  affine Dynkin
diagrams.

However,  the finite and the indefinite ones  give some QHE stringy
realizations. In particular,  one may distinguish between two models
depending on the values of the filling fraction. The first model
involves a quiver theory encoded in Dynkin  diagrams of   finite Lie
algebras.  This  leads to a fractional QHE with positive values.  It is best to give an example. Consider a particular vector charge like
$q_i=(1,\ldots,0)$. In this way, the above system reduces to a model
with the filling fraction  given by
\begin{equation}
\label{factor1} \nu(ADE)=K_{11}^{-1}.
\end{equation}
In the case of the finite Lie algebra $A_n$,  the  corresponding wrapped geometry consists
of $n$ intersecting 2-spheres as
 \be
    \mbox{
         \begin{picture}(20,30)(70,0)
        \unitlength=2cm
        \thicklines
    \put(0,0.2){\circle{.2}}
     \put(.1,0.2){\line(1,0){.5}}
     \put(.7,0.2){\circle{.2}}
     \put(.8,0.2){\line(1,0){.5}}
     \put(1.4,0.2){\circle{.2}}
     \put(1.6,0.2){$.\ .\ .\ .\ .\ .$}
     \put(2.5,0.2){\circle{.2}}
     \put(2.6,0.2){\line(1,0){.5}}
     \put(3.2,0.2){\circle{.2}}
     \put(-1.2,.15){$A_{n}:$}
  \end{picture}
} \label{ordAk}.\ee This  admits a
$T_{n,1,1}$ realization  and its Cartan matrix reads $K_{ij}$ reads as
\begin{equation}
K_{ij}=-\delta_{i-1,j}+2\delta_{i,j}-\delta_{i+1,j}.
\end{equation}
Using equations  (4.5) and (4.6), its determinant is given by
\begin{equation}
det\;K(T_{n,1,1})=n+1.
\end{equation}
From the matrix inverse, it is easy to see that the corresponding
 filling  fraction  reduces to
\begin{equation}
\label{factor1} \nu(A_n)=\frac{n}{n+1}.
\end{equation}
It should be interesting to note that  one can get filling fractions
with odd-denominator using a special choice of the singularities of
ALE spaces. In the case of the finite $A_n$ geometries, we  recover
some very known values observed experimentally including the Jain
series given by \be \nu^{\hbox{\tiny Jain}}=\frac{m}{2p m\pm 1}. \ee
Specializing the computation to the same vector charge  as before
and taking $n=2m$, we obtain $\nu=\frac{2m}{2m+1}$ which coincides
with two times the most stable part of the Jain's sequence with
$p=1$. The full series can
be recovered  exactly  if we replace the matrix
$K_{ij}$ that appears in the action (\ref{Chern})  by $2K_{ij}$.
In the algebraic side, this can be obtained
by fixing  the normalization of the inner product  to be such that
the highest weight of the adjoint representation has norm 1. It is
worth noting that other series  can be obtained in the context of
M-theory compactified on eight dimensional manifolds. This will be
given in future works.

The second model is described by  a quiver  gauge theory based on
indefinite geometries  and gives  QHE stringy realization with
negative filling factor. This negative value could have a physical
interpretation in  the physics of semi-conductors. To see that
consider a simple example of the indefinite Lie algebras given by
the
 hyperbolic symmetry  ($\widehat{HA}_2$). The derivation of this  hyperbolic geometry is based on the
same philosophy one uses in the building of the affine Dynkin
diagrams from the finite ones by adding a node. In other words, by
cutting this node in $\widehat{HA}_2$, the resulting sub-diagram
coincides with the affine Lie algebra $\widehat{A%
}_{2}$. The new Dynkin graph looks like

 \begin{equation} \mbox{
         \begin{picture}(20,90)(50,-20)
        \unitlength=2cm
        \thicklines
      \put(-1.2,.3){$\widehat{HA}_2:$}
\put(-0.7,0){\circle{.2}}
     \put(-0.6,0){\line(1,0){.5}}
    \put(0,0){\circle{.2}}
\put(0.1,0){\line(1,0){2.3}}
     \put(2.5,0){\circle{.2}}
     \put(0,.1){\line(2,1){1.15}}
     \put(1.25,.7){\circle{.2}}
     \put(2.5,.1){\line(-2,1){1.15}}
  \end{picture}
}  \label{affAk}
\end{equation}%
Its  Cartan  matrix takes the form
\begin{equation}
K_{ij}(\widehat{HA}_{2})=\left(
\begin{array}{ccccc}
2 & -1 & 0 & 0 &  \\
-1 & 2 & -1 & -1 &  \\
0 & -1 & 2 & -1 &  \\
0 & -1 & -1 & 2 &
\end{array}%
\right)
\end{equation}%
and its  inverse  is given by
\begin{equation}
K^{-1}_{ij}(\widehat{HA}_{2})= \left(
\begin{array}{cccc}
0 & -1 & -1 & -1 \\
-1 & -2 & -2 & -2 \\
-1 & -2 & -\frac{4}{3} & -\frac{5}{3} \\
-1 & -2 & -\frac{5}{3} & -\frac{4}{3}%
\end{array} \right).
\end{equation}
Now, using   equation (4.4) and taking a  vector charge like
$q_i=(0,1,0,0)$, the corresponding filling fraction  is
\begin{equation}
\label{factor1} \nu(\widehat{HA}_{2})=-2
\end{equation}
This value could be  related to some results in the physics of  graphene. The
latter  is a monoatomic layer of graphite with Carbon atoms arranged
in tow dimensional honeycomb lattice representation. The electronic
structure of the graphene can be modelled by fermions localized in
two  dimensions.   We expect that  the  above negative
value $\nu =-2$  can be related to holes in the graphene studied in
\cite{graphene}. The model with
 $\nu =2$ might  be related to it by particle-hole symmetry. It would therefore be of interest to try to extract more  information on this  string theory conection with graphene.   We believe that this
conection  deserves to be studied further.
\section{Discussions and open questions}
In this paper, we have  engineered   stringy realizations of the QHE
in 1+2 dimensions from the compactification of type  IIA superstring
on the K3 surface. Using  the results of the attractor horizon
geometries of the extremal black 2-branes on the Calabi-Yau
manifolds and based on the Hodge diamond of the K3 surface, we have
given  two different  realizations of this effect. In particular, we
have presented  hierarchical stringy  descriptions using quiver
gauge models living  on a stack of D4-branes wrapping intersecting
2-spheres which are  classified  by Cartan matrices of Lie algebras. These representations
could be related to the realization of  the QHE in 1+5 dimensions
involving unwrapped D4-branes in ten dimensions \cite{F}. This may
be done by wrapping D4-branes on the
 2-spheres  in the  compactification on the  K3 surface. \\ Using ideas borrowed from  the
classification of K3 surface singularities, we have studied the QHE
realizations  with filling fractions  depending on local type IIA
intersecting geometries.  As  results, we have found that there are
no QHE stringy descriptions    based on   type IIA
geometries corresponding  singularities classified by affine Kac-Moody Lie algebras.   The
stringy realizations of QHE  can be done only for the finite and
 indefinite  geometries. In particular, the finite Lie algebras are associated   with fractional filling factors; while
  the indefinite ones  classify    models with negative  filling
 factors, which could   be related to  the physics of   holes in the
 graphene.  It should be interesting  to develop this observation. This can go in the same sense
 of the string theory realization  of the graphene using  of D3-D7 brane system  studied recently in \cite{Rey}. It would
also be interesting to understand the link between our results and the ones in  \cite{BR} based on dyonic black holes.

An interesting question  for further investigations concerns  a
direct derivation of the  QHE in $1+2$  from    string theory
compactified  on seven real dimensional manifolds.  To make contact
with this  present work, a possible candidate could be  G2 manifolds
with K3-fibrations. On the other hand, the result of this paper
could be up lifted to M-theory. In this way,  the  QHE
configurations can be reproduced by wrapping M2 and M5-branes on
appropriate cycles in the K3 surface. It would be also worth
investigating a direct compactification on eight dimensional
manifolds. A  special interest might be devoted to incorporate the
compactification  on  Calabi-Yau fourfolds  or   Spin(7) manifolds
with K3-frabations.

 On the other hand, motivated by   results of
supersymmetric matrix model of the QHE given in
 \cite{GJSS}, it should be interesting to study realizations in terms  of
 local  K3  supersurface with $ADE$ geometries  studied in \cite{ABDS}.
  We hope,
 all these  open questions will be addressed in  future  work.

\vspace{0.5cm} \textbf{Acknowledgments.} This work has been
supported  CICYT (grant FPA2009-09638),  DGIID-DGA (grant
2007-E24/2) and  PCI-AECI (grant 07 and 10).  We thank M. Asorey, L. J.
Boya, I. Cavero-Pel\'aez,  M. P.  Garcia del Moral, J. M. Munoz-Castaneda,  E. H. Saidi, L. Segui for discussions and collaborations. AB   would like to thank S. J. Rey for sending him  the paper  \cite{Rey}. He would  also thank  Montanez-Naz  family for very  kind  hospitality.

\end{document}